\documentclass[10pt]{article}
\usepackage{paralist,amsmath,amsthm}
\usepackage[small,bf]{caption}
\usepackage{emp}

\renewcommand{\baselinestretch}{1.5}

\newtheorem{theorem}{Theorem}

\newcommand{\R}{\textbf{R}}
\newcommand{\Z}{\textbf{Z}}
\newcommand{\nix}[1]{}
\newcommand{\ie}{\textit{i.e.}}
\begin{document}
\begin{empfile}

\title{\vspace*{-15mm}\textbf{Scheduling Sensors by Tiling Lattices\footnote{
The research by A.K. was supported by NSF CAREER award CCF 0347310 and
NSF grant CCF 0622201.  The research by H.L. was supported in part by NSF
grant CNS-0614929. The
research by J.W. was supported in part by NSF grant 0500265 and Texas
Higher Education Coordinating Board grant ARP-00512-0007-2006.
}}
}
\author{Andreas Klappenecker, Hyunyoung Lee, and Jennifer L.~Welch}
\date{}
\maketitle 

\begin{abstract}
\noindent 
Suppose that wirelessly communicating sensors are placed in a regular
fashion on the points of a lattice. Common communication protocols
allow the sensors to broadcast messages at arbitrary times, which can
lead to problems should two sensors broadcast at the same time.  It is
shown that one can exploit a tiling of the lattice to derive a
deterministic periodic schedule for the broadcast communication of
sensors that is guaranteed to be collision-free.  The proposed
schedule is shown to be optimal in the number of time slots.
\end{abstract}


\section{Introduction}
Sensors are sometimes distributed in a regular fashion to monitor an
area. We assume that the sensors use wireless communication. Most
wireless communication protocols allow the sensors to send at
arbitrary times. However, this can cause the following \textit{collision 
problems}: 
If two distinct sensors $A$ and $B$ send at the same time and
$B$ is within the interference range of $A$, then 
frequently hardware limitations prevent $B$ from receiving  
the message of $A$ correctly. In addition,
if two distinct sensors $A$ and $B$ send at the same time and a
sensor $C$ is within interference range of both $A$ and $B$, then $C$ will
not be able to correctly receive either message.
In these cases, the sensors $A$ and $B$ need to
resend their messages, which is evidently a waste of energy.

Let us assume that the sensors have access to the current time,
represented by an integer $t$.  One can assign each sensor node an
integer $k$ and set up a periodic schedule such that a node with
integer $k$ is allowed to broadcast messages at time $t$ if and only
if $t\equiv k\pmod m$. The goal of this paper is to give a convenient
combinatorial formulation using lattice tilings that allows one to
assign optimal schedules with minimal number of time slots $m$ such
that no two sensors that are scheduled to broadcast simultaneously 
have intersecting interference ranges; we call such schedules
\textit{collision-free}.

\paragraph{Related Work.}
Since most communication protocols for wireless sensor networks are
probabilistic in nature, there exist few prior works that are directly
related to our approach. However, there exist a few notable exceptions
that we want to discuss here. 

Suppose for the moment that we are given a finite set of $k$ sensors
that share the same frequency band for communication. The simplest way
to ensure that the communication will be collision-free, is to use a
time division multiple access (TDMA) scheme. Here each of the $k$
sensors is assigned a different time slot and scheduling is done in a
round robin fashion. Because of its simplicity, this scheme is used in
many systems, see e.g.~\cite[Chapter 3.4]{schiller03}.  The obvious
disadvantage of TDMA is that it does not scale: If the number $k$ of
sensors is large, then the sensors cannot communicate frequently
enough.

The basic TDMA scheme does not take advantage of the fact that each sensor
typically affects only small number of neighboring sensors by its
radio communication. This prompts the question whether one can modify
the TDMA scheme and find a schedule with $m$ time slots that 
is collision-free.
To answer this
question, consider a directed graph that has a node for each sensor
and an edge from vertex $v$ to vertex $u$ if and only if $u$ is 
affected by the radio communication of $v$.  A valid schedule with $m$
time slots corresponds to a distance-2 coloring with $m$ colors, that
is, all vertices of distance $\le 2$ must be assigned a different
color (= time slot) to avoid collision problems. Therefore, the number
of time slots $m$ of an optimal collision-free schedule coincides with
the chromatic number of a distance-2 coloring. The distance-2 coloring
problem is also known as the broadcast scheduling problem in the
networking community.

McCormick has shown that the decision problem whether a given graph
has a distance-2 coloring with $m$ colors is
NP-complete~\cite{mccormick81}. Lloyd and Ramanathan showed that the
broadcast schedule problem even remains NP-complete when restricted to
planar graphs and $m=7$ time slots~\cite{lloyd92}.

Due to these intractability results, much of the subsequent research
focused on heuristics for finding optimal schedules; for instance,
Wang and Ansari used simulated annealing~\cite{wang97}, and Shi and
Wang used neural networks~\cite{shi05} to find optimal schedules.
Another popular direction of research are approximation algorithms for
broadcast scheduling algorithms, see e.g.~\cite{ramanathan92}.

\paragraph{Contributions.} 
The main contributions of this paper can be briefly summarized as
follows (the terminology is explained in the subsequent sections):
\begin{compactenum}[1)] 
\item We develop a method that allows one to derive an optimal
collision-free schedule from the tiling of a lattice.
\item Our scheme scales to an arbitrary number of sensors; in fact, we
formulate our schedules for an infinite number of sensors. Schedules
for a finite number of sensors are obtained by restriction, and these
schedules remain optimal under very mild conditions (given in the
conclusions). 
\item Our assumption on the set of prototiles ensures that an optimal
schedule is obtained regardless of the chosen tiling.  In Section 4,
we show that if our assumption on the set of prototiles is removed,
then in general one will not obtain an optimal schedule.
\end{compactenum}
We formulate our results for arbitrary lattices in arbitrary
dimensions, since the proofs are not more complicated than in the
familiar case of the two-dimensional square lattice. For the square
lattice, there are polynomial-time algorithms available to check
whether a given prototile can tile the lattice; thus, despite the fact
that finding optimal schedules is NP-hard in general, one can use our 
method to easily construct optimal schedules in the case of a single 
prototile.
This method of creating simple instances of an NP-hard problem might
be of independent interest.

\section{Lattice Tilings and Optimal Schedules} 
A Euclidean lattice $L$ is a discrete subgroup of $\R^d$ that spans
the Euclidean space $\R^d$ as a real vector space. In other words,
there exist $d$ vectors $\{v_1,\dots,v_d\}$ in $L$ that are linearly
independent over the real numbers such that
$$ L = \left\{ \sum_{k=1}^d a_kv_k\,\Bigg|\, a_k\in \Z \text{ for }
1\le k\le d\right\},$$ and for each vector $v$ in $L$ there exists an
open set containing $v$ but no other element of $L$.  In particular,
the group $L$ is isomorphic to the additive abelian group $\Z^d$. Two
examples of lattices in two dimensions are illustrated in
Figure~\ref{fig:lattice}.

\begin{figure}[htb]
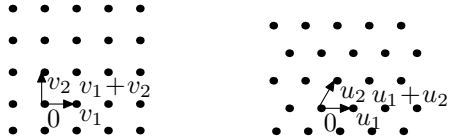


\begin{center}
\begin{emp}(40,40)
z1 = (1,0);
z2 = (0,1);
s = 12; 

drawarrow s*z1+s*z2 -- 2*s*z1+s*z2;
drawarrow s*z1+s*z2 -- s*z1+2*s*z2;
pickup pencircle scaled 3;

for k = 1 step s until 5*s: draw k*z1+1*z2 scaled 2; endfor;
for k = 1 step s until 5*s: draw k*z1+(s)*z2; endfor;
for k = 1 step s until 5*s: draw k*z1+(2*s)*z2; endfor;
for k = 1 step s until 5*s: draw k*z1+(3*s)*z2; endfor;
for k = 1 step s until 5*s: draw k*z1+(4*s)*z2; endfor;

label.lrt(btex $0$ etex,s*z1+s*z2);
label.lrt(btex $v_1$ etex,2*s*z1+s*z2);
label.lrt(btex $v_2$ etex,s*z1+2*s*z2);
label.lrt(btex $v_1\!+\!v_2$ etex,2*s*z1+2*s*z2);
\end{emp}
$\qquad\qquad$
\begin{emp}(40,40)
z1 = (1,0);
z2 = (0.5,0.866);

drawarrow s*z1+s*z2 -- 2*s*z1+s*z2;
drawarrow s*z1+s*z2 -- s*z1+2*s*z2;
pickup pencircle scaled 3;

for k = 1 step s until 5*s: draw k*z1+1*z2 scaled 2; endfor;
for k = 1 step s until 5*s: draw k*z1+(s)*z2; endfor;
for k = 1 step s until 5*s: draw (-s+k)*z1+(2*s)*z2; endfor;
for k = 1 step s until 5*s: draw (-s+k)*z1+(3*s)*z2; endfor;
for k = 1 step s until 5*s: draw (-2*s+k)*z1+(4*s)*z2; endfor;

label.lrt(btex $0$ etex,s*z1+s*z2);
label.lrt(btex $u_1$ etex,2*s*z1+s*z2);
label.lrt(btex $u_2$ etex,s*z1+2*s*z2);
label.lrt(btex $u_1\!+\!u_2$ etex,2*s*z1+2*s*z2);
\end{emp}
\end{center}

\caption{The figure on the left shows part of the square lattice $L_S=\Z^2$
that is generated by the vectors $v_1=(1,0)$ and $v_2=(0,1)$. The
figure on the right shows the hexagonal lattice $L_H$ that is
generated by the vectors $u_1=(1,0)$ and
$u_2=(\frac{1}{2},\frac{1}{2}\sqrt{3})$.}
\label{fig:lattice}
\end{figure}

Our goal is to find a deterministic collision-free periodic schedule
for sensors located at the points of a lattice $L$ that is
optimal in the number of time slots, \ie, no periodic schedule
with a shorter period can be found that is collision-free. 

We call a finite subset $N$ of $L$ a \textit{prototile} or a
\textit{neighborhood} of the point~$0$ if and only if it contains $0$
itself. The particular nature of $N$ will be determined for instance
by the type of antenna and by the signal strength used by the
sensor. The elements in $N$ are the sensors affected by wireless
communication of the sensor located at the point $0$ (that is,
only the elements in $N$ are within interference range of the
sensor located at the point $0$).  We will first assume a homogeneous
situation, namely the neighborhood affected by communication of the
sensor located at a point $t$ in $L$ is of the form $t+N = \{ t+n\,|\,
n\in N\},$ where the addition denotes the usual addition of vectors in
$\R^d$.  The set $t+N$ contains $t$, since $0$ is contained in $N$.
Some examples of neighborhoods $N$ are given in
Figure~\ref{fig:neighborhoods}.

\begin{figure}[htb]
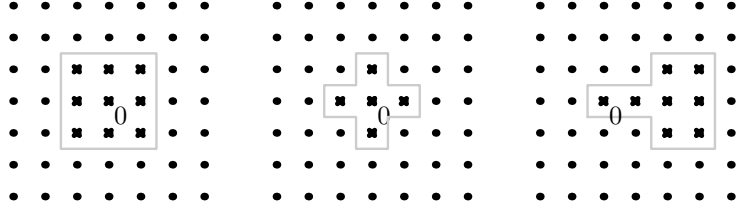

\begin{center}
\begin{emp}(50,50)
t = 12; 
z3 = t*(1,0);
z4 = t*(0,1);

def cross(expr a, b) =
begingroup
  pair g;
  g = a*z3+b*z4;
  pickup pencircle scaled 2;
  draw (g-(1,1))--(g+(1,1)); 
  draw (g-(1,-1))--(g+(1,-1));
endgroup;
enddef;

pickup pencircle scaled 3;
for k = 1 step 1 until 7: 
for l = 1 step 1 until 7:
  draw k*z3+l*z4;
endfor;
endfor;
label.lrt(btex 0 etex, 4*z3+4*z4);

for k = 3 upto 5: 
for l = 3 upto 5: 
  cross(k,l);
endfor;
endfor;

pickup pencircle scaled 1;
pair tg;
tg = 2.5*(z3+z4);
draw ((0,0)--(0,1)--(1,1)--(1,0)--cycle) scaled 36 shifted tg withcolor 0.8white;

\end{emp}
$\qquad$ % second neighborhood 
\begin{emp}(50,50)
z3 = t*(1,0);
z4 = t*(0,1);

def cross(expr a, b) =
begingroup
  pair g;
  g = a*z3+b*z4;
  pickup pencircle scaled 2;
  draw (g-(1,1))--(g+(1,1)); 
  draw (g-(1,-1))--(g+(1,-1));
endgroup;
enddef;

pickup pencircle scaled 3;
for k = 1 step 1 until 7: 
for l = 1 step 1 until 7:
  draw k*z3+l*z4;
endfor;
endfor;
label.lrt(btex 0 etex, 4*z3+4*z4);

for k = 3 upto 5: 
  cross(k,4); cross(4,k);
endfor;

pickup pencircle scaled 1;
draw ((0,1)--(0,2)--(1,2)--(1,3)--(2,3)--(2,2)--(3,2)--(3,1)--(2,1)--(2,0)--(1,0)--(1,1)--cycle) scaled (t) shifted (2.5*(z3+z4)) withcolor 0.8white;

\end{emp}
$\qquad$ % third neighborhood 
\begin{emp}(50,50)
z3 = t*(1,0);
z4 = t*(0,1);

def cross(expr a, b) =
begingroup
  pair g;
  g = a*z3+b*z4;
  pickup pencircle scaled 2;
  draw (g-(1,1))--(g+(1,1)); 
  draw (g-(1,-1))--(g+(1,-1));
endgroup;
enddef;

pickup pencircle scaled 3;
for k = 1 step 1 until 7: 
for l = 1 step 1 until 7:
  draw k*z3+l*z4;
endfor;
endfor;

for k = 5 upto 6: 
for l = 3 upto 5:
  cross(k,l);
endfor;
endfor;
cross(3,4); cross(4,4);
pickup pencircle scaled 1;
draw ((0,1)--(0,2)--(2,2)--(2,3)--(4,3)--(4,0)--(2,0)--(2,1)--(0,1)) scaled t shifted (2.5*(z3+z4)) withcolor 0.8white;
label.lrt(btex 0 etex, 3*z3+4*z4);
\end{emp}
\end{center}
\caption{The three figures illustrate some possible shapes of the
neighborhood $N$ of~0. The elements in $N$ are marked by small
crosses.  The left figure is a ball of radius 1 in the Chebycheff (or
$\ell_\infty$) metric. The figure in the middle is a ball of radius 1
in the Euclidean (or $\ell_2$) metric. The figure on the right
provides an example of a neighborhood where the sensor at 0 uses a
directional antenna.}\label{fig:neighborhoods}
\end{figure}

Our schedule will be a deterministic periodic schedule, that is,
each sensor is assigned a certain time slot and it is only allowed to
send during that time slot. Since our schedule is required to be free
of collision  problems, it follows that the sensors
located at distinct points $s$ and $t$ in $L$ cannot broadcast at the
same time unless
$$(s+N)\cap (t+N)=\emptyset.$$ 

Let $T$ denote a subset of $L$. We say that $T$ provides a
\textit{tiling} of $L$ with neighborhoods (or tiles) of the form $N$
if and only if the following two conditions hold:
\begin{compactenum}[\bf T1.]
\item $ \bigcup_{t\in T} (t+N) = T+N = L$,
\item $(s+N)\cap (t+N)=\emptyset$ for all distinct $s,t$ in $T$. 
\end{compactenum}
The set $T$ contains all the vectors that translate the prototile
$N$. Condition \textbf{T1} says that the whole lattice $L$ is 
covered by the translates $t+N$ of the prototile $N$, when $t$ ranges over
the elements of $T$. Condition \textbf{T2} simply says that the
translates of the tile $N$ do not overlap.

The tilings provide us with an elegant means to construct an optimal
deterministic schedule.

\begin{theorem}\label{th:thm1} 
Let $T$ be a tiling of a Euclidean lattice $L$ in $\R^d$ with
neighborhoods of the form $N$. Then there exists a deterministic
periodic schedule that avoids collision problems using $m=|N|$ time
slots. The schedule is optimal in the sense that one cannot achieve
this property with fewer than $m$ time slots.
\end{theorem}
\begin{proof}
Suppose that $N=\{n_1,\dots,n_m\}$ is the neighborhood of $0$.  For
$k$ in the range $1\le k\le m$, we schedule the sensors located at the
points $n_k+T$ at time $t\equiv k\pmod m$. We first notice that each
sensor located at a point in $L$ is scheduled at some point in time,
since $N+T=L$ by property \textbf{T1} of a tiling.

Seeking a contradiction, we assume that the schedule is not
collision-free. This means that at some time $k$ in the range $1\le
k\le m$ there exist sensors located at the positions $n_k+s$ and
$n_k+t$ with distinct $s$ and $t$ in $T$ such that $(n_k+s+N)\cap
(n_k+t+N)\neq \emptyset$.  However, this would imply that $(s+N)\cap
(t+N)\neq \emptyset$ for distinct $s$ and $t$ in $T$, contradicting
property \textbf{T2} of a tiling. It follows that our schedule is
collision-free.

It remains to prove the optimality of the schedule. Seeking a
contradiction, we assume that there exists a schedule with $m_0<m$
time slots that is collision-free.  This means that
for some time slot $k$ in the range $1\le k\le m_0$ two elements $n'$
and $n''$ of $N$ must be scheduled.  However, this would imply that
the element $n'+n''$ is contained in both sets $(n'+N)$ and $(n''+N)$,
contradicting the assumption that the schedule with $m_0$ time slots is
collision-free.
\end{proof}

We illustrate some aspects of the proof of the previous theorem in
Figure~\ref{fig:thm1a}. 

\begin{figure}[htb]
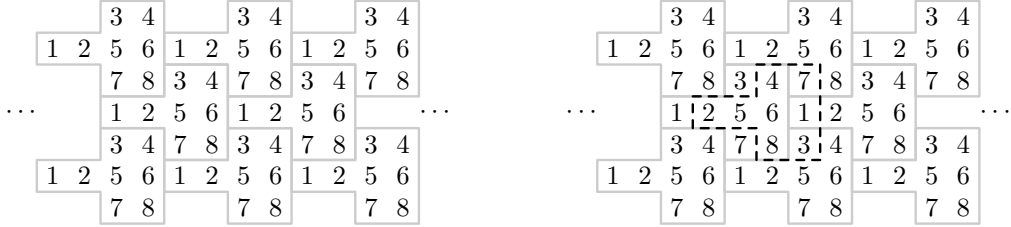

\begin{center}
\begin{emp}(50,50)
z3 = t*(1,0);
z4 = t*(0,1);

picture tile; 
vardef pnt(expr a,b) = (a*z3+b*z4) enddef;

pickup pencircle scaled 1;
path p; 
p = ((0,1)--(0,2)--(2,2)--(2,3)--(4,3)--(4,0)--(2,0)--(2,1)--(0,1)--cycle);
draw p scaled t shifted (0.5*z3+0.5*z4) withcolor 0.8white;

label(btex $1$ etex, pnt(1,2));
label(btex $2$ etex, pnt(2,2));
label(btex $3$ etex, pnt(3,3));
label(btex $4$ etex, pnt(4,3));
label(btex $5$ etex, pnt(3,2));
label(btex $6$ etex, pnt(4,2));
label(btex $7$ etex, pnt(3,1));
label(btex $8$ etex, pnt(4,1));

tile := currentpicture; 
currentpicture := nullpicture;

draw tile;
draw tile shifted pnt(2,2);
draw tile shifted pnt(0,4);
draw tile shifted pnt(4,0);
draw tile shifted pnt(4,4);
draw tile shifted pnt(6,2);
draw tile shifted pnt(8,0);
draw tile shifted pnt(8,4);

label(btex $\cdots$ etex, pnt(0,4));
label(btex $\cdots$ etex, pnt(13,4));
\end{emp}
$\qquad\qquad$
\begin{emp}(50,50)
z3 = t*(1,0);
z4 = t*(0,1);

picture tile;
vardef pnt(expr a,b) = (a*z3+b*z4) enddef;

pickup pencircle scaled 1;
path p;
p = ((0,1)--(0,2)--(2,2)--(2,3)--(4,3)--(4,0)--(2,0)--(2,1)--(0,1)--cycle);
draw p scaled t shifted (0.5*z3+0.5*z4) withcolor 0.8white;

label(btex $1$ etex, pnt(1,2));
label(btex $2$ etex, pnt(2,2));
label(btex $3$ etex, pnt(3,3));
label(btex $4$ etex, pnt(4,3));
label(btex $5$ etex, pnt(3,2));
label(btex $6$ etex, pnt(4,2));
label(btex $7$ etex, pnt(3,1));
label(btex $8$ etex, pnt(4,1));

tile := currentpicture;
currentpicture := nullpicture;

draw tile;
draw tile shifted pnt(2,2);
draw tile shifted pnt(0,4);
draw tile shifted pnt(4,0);
draw tile shifted pnt(4,4);
draw tile shifted pnt(6,2);
draw tile shifted pnt(8,0);
draw tile shifted pnt(8,4);

label(btex $\cdots$ etex, pnt(0,4));
label(btex $\cdots$ etex, pnt(13,4));

draw p scaled t shifted pnt(3.5,2.5) dashed evenly;
\end{emp}
\end{center}\caption{The figure on the left illustrates the previous theorem 
using a tiling with the neighborhood $N$ given by the rightmost
example of Figure~\ref{fig:neighborhoods}. Each of the eight elements
of $N$ is assigned a time slot from 1 to 8, the translated versions
$t+N$ with $t$ in $T$ have the time slots at the corresponding
translated positions. As a result, the broadcast during time slot 1
affects only the neighborhoods shown in the tiling depicted by the
figure.
The figure on the right shows one dashed neighborhood of a sensor
broadcasting during time slot 2. Considering the neighborhoods of all
sensors broadcasting during time slot 2 one obtains once again a
tiling, namely the tiling of $L_S$ obtained by right shifting the
(solid gray) neighborhoods of the sensors broadcasting during time step~1. 
} \label{fig:thm1a} 
\end{figure}

\section{Existence of Tilings} 
Our concept of tiling a lattice with translates of a prototile $N$
turned out to be convenient for our purposes. In this section, we
relate the tilings of a lattice to tilings of the Euclidean space
$\R^d$, so that we can benefit from the large number of results that
are available in the literature.

Any tiling of a lattice $L$ can be converted into a tiling of $\R^d$
as follows. Let $K$ denote the union of the closed Voronoi regions
about the points in $N$. Then the translates $t+K$ with $t$ in $T$
yield a tiling of $\R^d$. Conversely, any tiling of $\R^d$ with
translates of a tile consisting of the union of Voronoi regions of
points in $L$ evidently yields a tiling of the lattice $L$ in our
sense. Figure~\ref{fig:quasipolyominoes} shows some two-dimensional
examples of Voronoi regions.

\begin{figure}[htb]
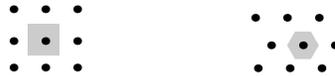

\begin{center}
\begin{emp}(40,40)
z1 = (1,0);
z2 = (0,1);

pickup pencircle scaled 1;
fill ((0,0)--(1,0)--(1,1)--(0,1)--cycle) 
scaled s shifted (6.4*(z1+z2)) withcolor 0.8white;

pickup pencircle scaled 3;

for k = 1 step s until 3*s: draw k*z1+1*z2 scaled 2; endfor;
for k = 1 step s until 3*s: draw k*z1+(s)*z2; endfor;
for k = 1 step s until 3*s: draw k*z1+(2*s)*z2; endfor;

\end{emp}
$\qquad\qquad\qquad$
\begin{emp}(40,40)
z1 = (1,0);
z2 = (0.5,0.866);

vardef pnt(expr a,b) = (a*z1+b*z2) enddef;

pickup pencircle scaled 1;
fill (pnt(0,0)--pnt(1,0)--pnt(1,1)--pnt(0,2)
--pnt(-1,2)--pnt(-1,1)--pnt(0,0)--cycle) 
scaled (0.5*s) shifted (z1+pnt(1*s,0.5*s)) withcolor 0.8white; 

pickup pencircle scaled 3;

for k = 1 step s until 3*s: draw k*z1+1*z2 scaled 2; endfor;
for k = 1 step s until 3*s: draw k*z1+(s)*z2; endfor;
for k = 1 step s until 3*s: draw (-s+k)*z1+(2*s)*z2; endfor;

\end{emp}
\end{center}

\caption{(a) The figure on the left shows that the Voronoi region
about a point in the square lattice $L_S$ is given by a square of unit
length; a tile $K$ in the plane obtained by a union of unit squares
about points in $L_S$ is called a quasi-polyomino. (b) A Voronoi
region about a point in the hexagonal lattice $L_H$ is a hexagon; a
tile $K$ in the plane obtained from a union of these Voronoi-hexagons
about points in $L_H$ is called a quasi-polyhex.}
\label{fig:quasipolyominoes} 
\end{figure}

The union of Voronoi regions about points in a lattice are also known
as quasi-polyforms. A quasi-polyform that is homeomorphic to the unit
ball in $\R^d$ is known as a polyform. The books by Gr\"unbaum and
Shepherd~\cite{gruenbaum87} and by Stein and Szab\'o~\cite{stein94}
contain numerous examples of tilings obtained by translating quasi-polyforms
(and especially polyforms). The polyforms in the square grid
$L_S=\Z^2$ are called polyominoes, the most well-known type of
polyforms; see Golomb's book~\cite{golomb94}. By abuse of language, we
will also refer to a prototile $N$ in $L_S$ as a polyomino if the
union of the Voronoi regions of $N$ form a polyomino.

A prototile $N$ in a lattice $L$ that admits a tiling is called
\textit{exact}. It is natural to ask the following question: 
\begin{compactenum}[\bf Q1.]
\item When is a given prototile $N$ exact, \ie, when does there exist
a subset $T$ of $L$ such that the conditions \textbf{T1} and
\textbf{T2} are satisfied?
\end{compactenum}
Beauquier and Nivat gave a simple criterion that allows one to answer
\textbf{Q1} for polynominos in the square lattice $L_S$. Roughly
speaking, their criterion says that if $N$ can be surrounded by
translates of itself such that there are no gaps or holes, then $N$ is
exact; see~\cite{beauquier91} for details.  In particular, it
immediately follows that each prototile shown in
Figure~\ref{fig:neighborhoods} is exact.

Algorithmic criteria for deciding the question \textbf{Q1} are
particularly interesting.  For polyominoes in the square lattice
$L_S$, one can decide this question in time polynomial in the length
of the boundary of the polyomino (described by a word over the
alphabet $\{u,d,l,r\}$, which is short for up, down, left, and right),
as Wijshoff and van Leeuwen have shown~\cite{wijshoff84}. The
characterization of exactness of a polyomino by Beauquier and Nivat
\cite{beauquier91} mentioned above leads to an $O(n^4)$ algorithm,
where $n$ is the length of the word describing the boundary. Recently,
Gambini and Vuillon~\cite{gambini07} derived an improved $O(n^2)$
algorithm for this problem.

Less is known for arbitrary (not necessarily connected) prototiles in
a general lattice. Szegedy~\cite{szegedy98} derived an algorithm to
decide whether a prototile $N$ in a lattice $L$ is exact assuming that
the cardinality of $N$ is a prime or is equal to 4.

\section{Generalization to Several Prototiles} 
We have seen that the conditions for tiling a lattice with a single
prototile are somewhat restrictive. For example, we might want to
allow different rotated versions of the tile if the radiation pattern
of the antenna used by a sensor is asymmetrical. We might want to
consider different tiles corresponding to various different signal
strength settings. Furthermore, we might want to allow sensors with
various different styles of antenna.

We can accommodate all these different situations by allowing
translates of several prototiles instead of just a single one.  In
this section, we show that one can still obtain an optimal periodic
schedule which guarantees that the schedule is collision-free, as long
as sensors of the same type and setting are deployed within each tile
and a constraint on the tiles is satisfied.

Let $L$ be a lattice in $\R^d$. Let $N_1,\dots,N_n$ be prototiles in
the lattice $L$, that is, $N_k$ is a subset of $L$ that contains $0$ for 
$1\le k\le n$. Let $T_1,\dots,T_n$ be pairwise disjoint
nonempty subsets of $L$. We say that $T_1,\dots, T_n$ provide a tiling
of $L$ with prototiles $N_1,\dots, N_n$ if and
only if the following two conditions are satisfied:
\begin{compactenum}[\bf GT1.]
\item $\displaystyle \bigcup_{k=1}^n \bigcup_{t_k\in T_k}
(t_k+N_k)=\bigcup_{k=1}^n (T_k+N_k)= L.$
\item For all $k,\ell\in \{1,\dots,n\}$, we have $(s_k+N_k)\cap
(t_\ell+N_\ell)=\emptyset$ for all $s_k$ in $T_k$ and $t_\ell$ in
$T_\ell$ such that $s_k\neq t_\ell$.
\end{compactenum}
\smallskip

\noindent 
Condition \textbf{GT1} ensures that the lattice $L$ is covered by
translates of the prototiles $N_1,\dots,N_n$. Condition \textbf{GT2}
ensures that two distinct tiles will not overlap.  The set $T_k$
contains all vectors that are used to translate the tile $N_k$, that
is, the set $\{t_k+N_k\,|\, t_k \in T_k\}$ contains all shifted
versions of $N_k$ that occur in the tiling of $L$. Since the sets
$T_1,\dots, T_n$ are pairwise disjoint, it is clear that
$(s_k+N_k)\cap (t_\ell+N_\ell)=\emptyset$ whenever $k\neq \ell$.
Condition \textbf{GT2} requires further that the translates of the
prototile $N_k$ with elements in $T_k$ do not overlap. 

We will call a tiling of $L$ \textit{respectable} if and only if the
prototile $N_1$ contains all other prototiles $N_k$, that is,
$N_1\supseteq N_k$ for $2\le k\le n$. If this is the case, then we
call $N_1$ the \textit{respectable prototile}.

Suppose that we are given a tiling $T_1,\dots,T_n$ of $L$ respectively
with neighborhoods of the form $N_1,\dots, N_n$.  We will assume that
the sensors are deployed in the following fashion:  
\begin{compactenum}[\bf D1.] 
\item A sensor at location $s_k$ in the neighborhood $t_k+N_k$ of an
element $t_k$ in $T_k$ affects precisely the neighbors $s_k+N_k$ by
interference, where $k$ is in the range $1\le k\le n$.
\end{compactenum}
Loosely speaking, condition \textbf{D1} says that all elements in the
neighborhood $t_k+N_k$ have neighborhood type $N_k$.

\begin{theorem} \label{th:thm2} 
Let $T_1,\dots,T_n$ be a respectable tiling of a Euclidean lattice $L$
with neighborhoods of the type $N_1,\dots,N_n$. Suppose that the
sensors are deployed according to the
scheme~\textbf{D1}.  Then there exists a deterministic periodic
schedule that avoids collision problems using
$m=|N_1|$ time slots. The schedule is optimal in the sense that one
cannot achieve this property with fewer than $m$ time slots. 
\end{theorem}
\begin{proof}
The periodic schedule is specified as follows. 
Let $N=\bigcup_{k=1}^n
N_k=\{n_1,\dots,n_m\}$. 
For all $\ell$ in the range $1\le \ell\le n$, 
we schedule the elements $n_k+T_\ell$ at time
$t\equiv k\pmod m$ if and only if $n_k$ is contained in the
neighborhood $N_\ell$. 

Notice that all elements in $L$ will be scheduled at some point in
time by property \textbf{GT1}. Furthermore, condition \textbf{GT2}
 ensures that an element in $L$ is not scheduled more
than once within $m$ consecutive time steps.

We claim that this schedule is collision-free.  Seeking a
contradiction, we assume that two distinct elements in $L$ are
scheduled at the same time, but yield a collision problem. In other
words, there must exist integers $k$ and $\ell$ in the range $1\le
k,\ell\le n$, an element $n\in N$ such that $n$ is contained in both
$N_k$ and $N_\ell$, and elements $s_k\in T_k$ and $t_\ell\in T_\ell$
with $s_k\neq t_\ell$ such that $(n+s_k+N_k) \cap
(n+t_\ell+N_\ell)\neq \emptyset$. This implies that $(s_k+N_k)\cap
(t_\ell +N_\ell)\neq \emptyset$ for $s_k\neq t_\ell$, contradicting
property \textbf{GT2}. Therefore, our schedule is collision-free.

Without loss of generality, we may assume that the point $0$ in $L$
has a respectable neighborhood $N_1$ (otherwise, simply shift the
tiling such that this condition is satisfied). Seeking a
contradiction, we assume that there exists a deterministic periodic
schedule with $m_0<m$ time slots that is collision-free. It follows that there must exist two distinct elements $n'$ and
$n''$ in $N_1$ that are scheduled at the same time. However, this
would imply that the element $n'+n''$ is contained in both $n'+N_1$
and $n''+N_1$; thus, $(n'+N_1)\cap (n''+N_1)\neq \emptyset$,
contradicting the fact that the schedule with $m_0$ time slots is
collision-free.
\end{proof}

The previous theorem is a natural generalization of
Theorem~\ref{th:thm1}. A salient feature of Theorems~\ref{th:thm1}
and~\ref{th:thm2} is that the optimal schedule is independent of the
nature of the tiling of $L$. 

Notice that one can obtain a collision-free periodic schedule even
when there does not exist a respectable prototile. In fact, the
respectable prototile was only used in the last part of the proof of
Theorem~\ref{th:thm2} to establish the optimality of the
schedule. Therefore, one might wonder what will happen in the
non-respectable case.

Let us agree on some ground rules. We would like to maintain the fact
that for each translated version of a prototile the schedule is the
same, as this simplifies configuring the sensor network. However, in
the non-respectable case we might have different prototiles of the
same size, so we allow that the schedules in the different prototiles
can be independently chosen, as long as this does not lead to collision problems. Figure~\ref{fig:non-respectable}
shows that the number of time steps in an optimal schedule depends on
the chosen tiling when the tiling is non-respectable.

\begin{figure}[htb]
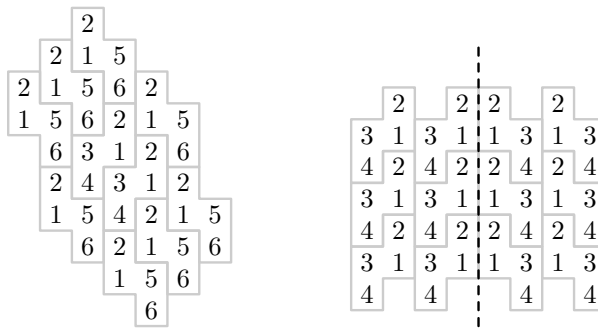

\begin{center}
\begin{emp}(50,50)
z3 = t*(1,0);
z4 = t*(0,1);

picture ltile, rtile; 
vardef pnt(expr a,b) = (a*z3+b*z4) enddef;

pickup pencircle scaled 1;
path p; 
p = ((0,1)--(0,3)--(1,3)--(1,2)--(2,2)--(2,0)--(1,0)--(1,1)--(0,1)--cycle);
draw p scaled t shifted (0.5*z3+0.5*z4) withcolor 0.8white;

label(btex $1$ etex, pnt(1,2));
label(btex $2$ etex, pnt(1,3));
label(btex $5$ etex, pnt(2,2));
label(btex $6$ etex, pnt(2,1));

ltile := currentpicture; 
currentpicture := nullpicture;

path q; 
q = p reflectedabout ((2,0),(2,1)); 
draw q scaled t shifted (-1.5*z3+0.5*z4) withcolor 0.8white;

label(btex $4$ etex, pnt(1,1));
label(btex $3$ etex, pnt(1,2));
label(btex $1$ etex, pnt(2,2));
label(btex $2$ etex, pnt(2,3));

rtile := currentpicture; 
currentpicture := nullpicture;

draw ltile shifted pnt(0,5);
draw ltile shifted pnt(1,2);
draw ltile shifted pnt(3,0);
draw ltile shifted pnt(4,1);
draw ltile shifted pnt(5,2);
draw ltile shifted pnt(4,5);
draw ltile shifted pnt(2,7);
draw ltile shifted pnt(1,6);

draw rtile shifted pnt(3,3);
draw rtile shifted pnt(2,4);
\end{emp}
$\qquad\qquad$ 
\begin{emp}(50,50)
z3 = t*(1,0);
z4 = t*(0,1);

picture ltile, rtile; 
vardef pnt(expr a,b) = (a*z3+b*z4) enddef;

pickup pencircle scaled 1;
path p; 
p = ((0,1)--(0,3)--(1,3)--(1,2)--(2,2)--(2,0)--(1,0)--(1,1)--(0,1)--cycle);
draw p scaled t shifted (0.5*z3+0.5*z4) withcolor 0.8white;

label(btex $1$ etex, pnt(1,2));
label(btex $2$ etex, pnt(1,3));
label(btex $3$ etex, pnt(2,2));
label(btex $4$ etex, pnt(2,1));

ltile := currentpicture; 
currentpicture := nullpicture;

path q; 
q = p reflectedabout ((2,0),(2,1)); 
draw q scaled t shifted (-1.5*z3+0.5*z4) withcolor 0.8white;

label(btex $4$ etex, pnt(1,1));
label(btex $3$ etex, pnt(1,2));
label(btex $1$ etex, pnt(2,2));
label(btex $2$ etex, pnt(2,3));

rtile := currentpicture; 
currentpicture := nullpicture;

for j = 0 step 2 until 4: 
for i = 0 step 2 until 2:
  draw rtile shifted pnt(i,j);
endfor;
endfor;

for j = 0 step 2 until 4: 
for i = 4 step 2 until 6:
  draw ltile shifted pnt(i,j);
endfor;
endfor;

draw pnt(4.5,0)--pnt(4.5,9) dashed evenly;

\end{emp}
\end{center}

\caption{ The figure on the left shows a schedule for a
non-respectable tiling with two tetrominos (\ie, prototiles with 4
elements). The tiling contains two $Z$-shaped tetrominos that are
surrounded by $S$-shaped tetrominos (rotating the figure clockwise by
$90^\circ$ might help identifying the $S$ and $Z$ shapes). The
schedule was determined with the algorithm given in the proof of
Theorem~\ref{th:thm2}. It is not difficult (though tedious) to show
that this schedule with $m=6$ time steps is optimal.  However, if the
lattice is tiled in the symmetric fashion shown in the right figure,
then the optimal schedule has $m=4$ time steps. Therefore, in the
non-respectable case the number of time steps of an optimal schedule
depends on the chosen tiling.}\label{fig:non-respectable}
\end{figure}

\section{Conclusions} 
We have introduced a deterministic periodic schedule for sensors
using wireless communication that are placed on the points of a
lattice. We have shown that the schedule is optimal assuming that
there exists a respectable prototile. A natural question is whether the
schedule remains optimal if one restricts the schedule from the
lattice $L$ to a finite subset $D$ of $L$. This question has an
affirmative answer if $D$ contains a translate of the set $N_1+N_1$,
as the latter set consists of the respectable prototile $N_1$ and its
neighbors, in which case our optimality proof carries over without
change.

Another natural question is whether one can extend the method to the
case of mobile sensors. This question has an affirmative
answer. Indeed, one straightforward way is to use our schedule to
assign time slots to the locations rather than to the sensors. Let us
assume that the lattice points are spaced fine enough to ensure that
only one sensor is within a Voronoi region of a lattice point. If the
time slot $k$ is assigned to a lattice point $p$, then a sensor $s$
within the open Voronoi region about $p$ can send at time $t$ if and
only if $t\equiv k\pmod m$ and the interference range of $s$ fits
within the tile of $p$. Clearly, this yields a collision-free schedule
for mobile sensors. However, it should be stressed that there are many
other solutions possible, but a comparison of such methods is beyond
the scope of this paper.

\end{empfile} 

\renewcommand{\baselinestretch}{1.2}
\small\normalsize
%\bibliographystyle{plain}
%\bibliography{sensor-schedules} 

\begin{thebibliography}{10}

\bibitem{beauquier91}
D.~Beauquier and M.~Nivat.
\newblock On translating one polyomino to tile the plane.
\newblock {\em Discrete Comput. Geom.}, 6(6):575--592, 1991.

\bibitem{gambini07}
I.~Gambini and L.~Vuillon.
\newblock An algorithm for deciding if a polyomino tiles the plane.
\newblock {\em Theor. Inform. Appl.}, 41(2):147--155, 2007.

\bibitem{golomb94}
S.W. Golomb.
\newblock {\em Polyominoes}.
\newblock Princeton University Press, Princeton, NJ, second edition, 1994.

\bibitem{gruenbaum87}
B.~Gr{\"u}nbaum and G.C. Shephard.
\newblock {\em Tilings and Patterns}.
\newblock W.H. Freeman and Company, 1987.

\bibitem{lloyd92}
E.L. Lloyd and S.~Ramanathan.
\newblock On the complexity of distance-2 coloring.
\newblock In {\em Fourth Intl.\ Conf.\ on Computing and Information}, pages
  71--74, 1992.

\bibitem{mccormick81}
S.T. McCormick.
\newblock Optimal approximation of sparse {H}essians and its equivalence to a
  graph coloring problem.
\newblock Technical report SOL 81-22, Department of Operations Research,
  Stanford University, 1981.

\bibitem{ramanathan92}
S.~Ramanathan and Errol~L. Lloyd.
\newblock Scheduling algorithms for multi-hop radio networks.
\newblock {\em ACM SIGCOMM Computer Communication Review}, 22(4):211--222,
  1992.

\bibitem{schiller03}
J.~Schiller.
\newblock {\em Mobile Communications}.
\newblock Addison Wesley, 2nd edition, 2003.

\bibitem{shi05}
H.~Shi and L.~Wang.
\newblock Broadcast scheduling in wireless multihop networks using a
  neural-network-based hybrid algorithm.
\newblock {\em Neural Netw.}, 18(5-6):765--771, 2005.

\bibitem{stein94}
S.K. Stein and S.~Szab{\'o}.
\newblock {\em Algebra and Tiling -- Homomorphism in the service of Geometry}.
\newblock Number~25 in The Carus Mathematical Monographs. The Mathematical
  Association of America, 1994.

\bibitem{szegedy98}
M.~Szegedy.
\newblock Algorithms to tile the infinite grid with finite clusters.
\newblock In {\em 39th Annual Symposium on Foundations of Computer Science
  (FOCS '98)}, pages 137--147, 1998.

\bibitem{wang97}
G.~Wang and N.~Ansari.
\newblock Optimal broadcast scheduling in packet radio networks using mean
  field annealing.
\newblock {\em IEEE J. Selected Areas in Communications}, 15(2):250--260, 1997.

\bibitem{wijshoff84}
H.A.G. Wijshoff and J.~van Leeuwen.
\newblock Arbitrary versus periodic storage schemes and tessellations of the
  plane using one type of polyomino.
\newblock {\em Inform. and Control}, 62(1):1--25, 1984.

\end{thebibliography}

\end{document}